# Application of Long Short-Term Memory Recurrent Neural Networks Based on the BAT-MCS for Binary-State Network Approximated Time-Dependent Reliability Problems

Wei-Chang Yeh
Department of Industrial Engineering and Engineering Management
National Tsing Hua University
P.O. Box 24-60, Hsinchu, Taiwan 300, R.O.C.
yeh@ieee.org

*Abstract:* Reliability is an important tool for evaluating the performance of modern networks. Currently, it is NP-hard and #P-hard to calculate the exact reliability of a binary-state network when the reliability of each component is assumed to be fixed. However, this assumption is unrealistic because the reliability of each component always varies with time. To meet this practical requirement, we propose a new algorithm called the LSTM-BAT-MCS, based on long short-term memory (LSTM), the Monte Carlo simulation (MCS), and the binary-adaption-tree algorithm (BAT). The superiority of the proposed LSTM-BAT-MCS was demonstrated by experimental results of three benchmark networks with at most $10^{-4}$ mean square error.

## 1. INTRODUCTION

Networks are widely used in numerous applications including network resilience [1], wireless sensor networks [2], pipe networks [3], Internet of Things [4], communication networks [5], and social networks [6]. To evaluate and manage these networks, their structures can be modeled as systems in which each component (arcs and/or nodes) is in a binary state: either working or failed [7, 8]. As the binary-state network model is fundamental to all types of networks, it is necessary to evaluate the performance of binary-state networks.

Reliability is a common performance and functionality metric for networks. For a given value of $t$, reliability is expressed as follows:

$$R(t) = \Pr\{t \le T\}, \tag{1}$$

where $0 \le T$ is the time to system failure, $0 \le R(t)$, $R(0) = 1$, and $\lim_{t\to\infty} R(t) = 0$. Thus, reliability can be defined as the probability that an item will perform its intended functions within a specific time interval under the stated conditions [9].

Several algorithms have been proposed for binary-state network reliability problems [30, 31]. Depending on whether minimal paths (MPs) or minimal cuts (MCs) are required, these algorithms can be divided into four main categories:

1. Direct exact-reliability algorithms, such as the binary addition tree algorithm (BAT) [11], binary-decision diagram [12, 13, 14], and state-space algorithm [15], can calculate exact reliability directly. Generally, direct algorithms are more efficient than indirect ones.
2. Indirect exact-reliability algorithms are either entirely MP-based [16, 17, 18] or entirely MC-based [19, 20, 21]. Thus, the inclusion-exclusion method or sum-of-disjoint method is applied to calculate the exact reliability of the binary-state networks based on MPs or MCs. Note that the above procedures are both NP-hard and #P-hard [7, 8].
3. Direct approximated-reliability algorithms include Monte Carlo simulations (MCSs) [22], MCS-based algorithms [23, 24, 25, 26], and AppBAT [27]. These algorithms can overcome NP-hard and #P-hard obstacles to achieve good approximated reliability.
4. Indirect approximated-reliability algorithms, such as MP-based and MC-based bound algorithms [9, 28], focus on finding the bounds of reliability. While these algorithms are similar to indirect exact reliability algorithms, they require a smaller number of MPs or MCs to yield accurate results.

In all aforementioned reliability algorithms, it is always assumed that the reliability of each component is constant [6, 7, 15, 29] (for example, $R(t) = R$ for all $t$ in Eq.(1)). Thus, the time factor is ignored in the network reliability problem.

Unfortunately, as the reliability of real-life network components decreases with time, it is impractical to evaluate network performance without considering time as a variable. As practical networks experience fewer limitations [14, 30], they also grow in scale. Therefore, there is a need to solve the traditional network reliability algorithm while considering time differences.

Among exact-reliability algorithms, the BAT is more efficient than algorithms based on traditional search methods, such as breadth-first search (BFS) [16, 31], depth-first search (DFS) [32, 33], and universal generating function methodology (UGFM) [34]. Moreover, the BAT is very simple to code and flexible to customize without requiring extensive coding expertise using massive data structures [11, 13]. However, due to the NP-hard characteristic of the network reliability problem, exact-reliability algorithms are not suitable for adequately sized, constantly growing modern networks.

The MCS is straightforward and convenient for overcoming the NP-hard obstacle in assessing the reliability of different types of networks [22, 24, 25, 27, 28]. In general, MCS-based algorithms uniformly generate random values for nodes/arcs to simulate the state of the entire network repeatedly, and then average the results to approximate reliability. However, MCS-based algorithms cannot provide fixed and acceptable approximations of reliability even after generating a large number of random values.

The (self-adaptive) BAT-MCS was proposed by Yeh to improve the accuracy of the MCS and decrease the number of random values. Compared to the traditional MCS, the BAT-MCS is simple, efficient, and robust in terms of time complexity, solution quality, and variance, as shown experimentally [27].

Based on the above discussion, many algorithms have been proposed for binary-state network problems. However, none of them account for the decrease in component reliability over time. This decrease in reliability is the main problem shared by all reliability algorithms, and a new algorithm needs to be developed to solve it.

Deep learning is a type of machine learning in artificial intelligence based on artificial neural networks with many layers to progressively extract higher-level features from input data and build a composite representation [35, 36, 60, 61]. Since the 2010s, increasingly efficient methods for deep learning have been developed that benefit from recent improvements in both machine learning and computer hardware [62]. Currently, deep learning delivers the best solutions for various applications in image recognition [37], speech recognition [38], and natural language processing [39].

Long short-term memory (LSTM) [40] is a class of deep learning with an artificial recurrent neural network (RNN) architecture [41] that stores information over extended time intervals. Our study aimed to develop a new algorithm called the LSTM-BAT-MCS to calculate accurate, robust, and time-dependent estimators by integrating the advantage of LSTM in dealing with time-series data and the benefit of the BAT-MCS in providing accurate and robust approximations of reliability for binary-state networks.

The remainder of this paper is organized as follows. Section 2 introduces the acronyms, notations, nomenclature, and assumptions used in this study. Section 3 reviews the BAT, BAT-MCS, and LSTM. Section 4 discusses the proposed LSTM-BAT-MCS. Section 5 presents a computational experiment of the proposed LSTM-BAT-MCS performed on three large-scale networks, each with its own special time-dependent arc reliability function. Finally, Section 6 concludes the study.

## 2. ACRONYMS, NOTATIONS, NOMENCLATURE, AND ASSUMPTIONS

Acronyms, notations, assumptions, and nomenclatures are required for the proposed LSTM-BAT-MCS are presented in this section.

### 2.1 Acronyms

BAT: Binary-Addition-Tree Algorithm

MCS: Monte Carlo simulation

BAT-MCS the MCS based on BAT with self-adaptive simulation-number algorithm

LSA: Layered Search Algorithm

PLSA: Path-based LSA

ANN: Artificial neural network

RNN: recurrent ANN

LSTM: Long short-term memory

### 2.2 Notations

$|\bullet|$: number of elements in $\bullet$

||•||: number of coordinates in •

$T$ $T = \{1, 2, \ldots, \tau\}$ is the time step set

Pr($t$, •): probability of • at time step $t \in T$ and Pr($t$, •) = Pr(•) if time step can be ignored

$a_i$: arc $i$

$V$: $V = \{1, 2, \ldots, n\}$ is the node set

$E$: $E = \{a_1, a_2, \ldots, a_m\}$ is the arc set

$X$: (arc) state vector

$X(a_i)$: value of the $a_i$ in $X$ for $i = 1, 2, \ldots, m$, e.g., $X(a_1) = X(a_2) = X(a_3) = 1$ and $X(a_4) = X(a_5) = 0$ if $X = (1, 1, 1, 0, 0)$.

**D**: time-dependent (arc) state distribution listed $\mathbf{D}(t, a) = \Pr(t, a)$ for all $a \in E$ at time step $t$, e.g., **D** is listed in Table 1.

**Table 1.** Time-dependent state distribution **D**

| $i$ | $\mathbf{D}(0, a_i)$ | $\mathbf{D}(1, a_i)$ | $\mathbf{D}(2, a_i)$ |
|---|---|---|---|
| 1 | 0.90 | 0.80 | 0.70 |
| 2 | 0.80 | 0.70 | 0.75 |
| 3 | 0.70 | 0.65 | 0.60 |
| 4 | 0.60 | 0.58 | 0.56 |
| 5 | 0.50 | 0.45 | 0.35 |

$G(V, E)$: an undirected graph with $V$ and $E$. For example, Figure 1 is a graph with $V = \{1, 2, 3, 4\}$, $E = \{a_1, a_2, a_3, a_4, a_5\}$, source node 1, and sink node 4

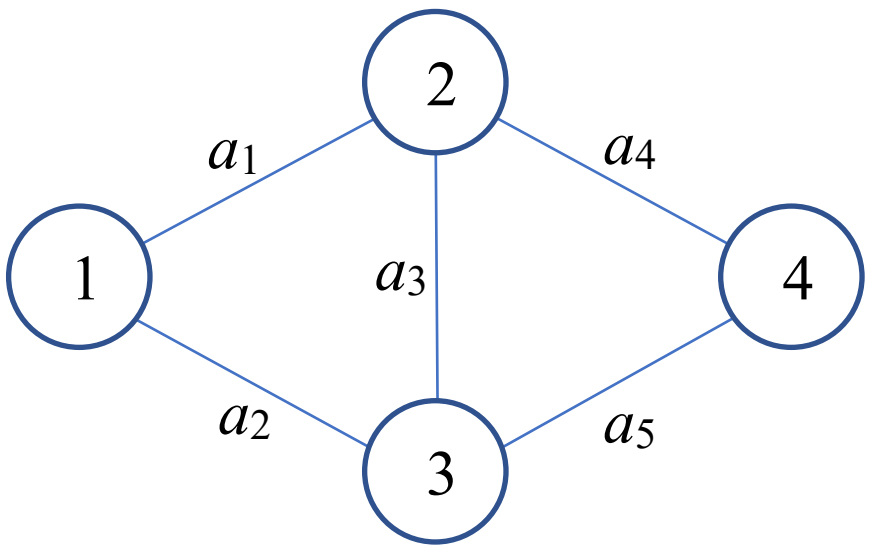

**Figure 1.** Example graph

$G(V, E, \mathbf{D})$: binary-state network constituted by $G(V, E)$ and **D**. For example, Figure 1 is a binary-state network if **D** is as given in Table 1.

$G(X)$: subgraph corresponding to $X$ if $G(X) = G(V, \{a \in E \mid \text{for all } a \text{ with } X(a) = 1\}, \mathbf{D})$. For example, $G(X)$ is depicted in Figure 2, where $X = (1, 1, 1, 0, 0)$ in Figure 1.

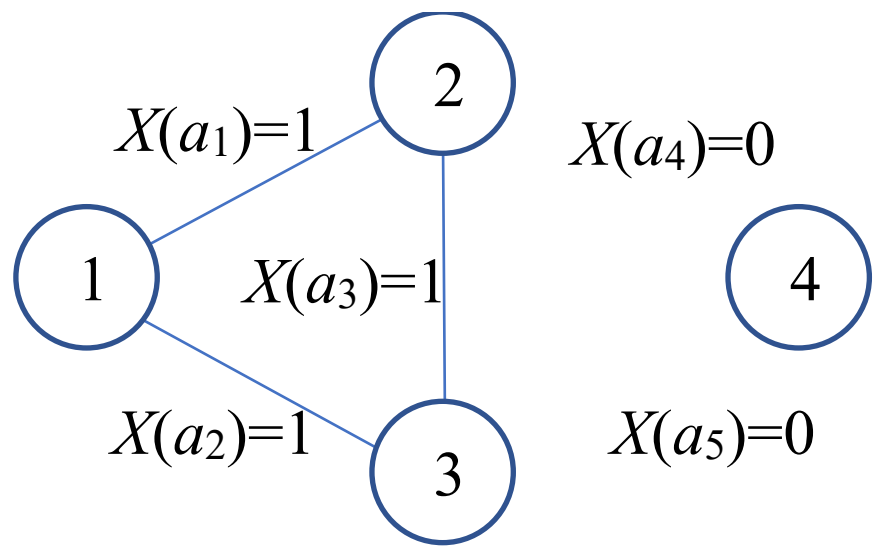

**Figure 2.** $X = (1, 1, 1, 0, 0)$ and $G(X)$, where $G(V, E)$ is shown in Figure 1.

$R(t)$: exact reliability at time step $t$

$R^*(\bullet)$: approximated reliability of • obtained from the BAT-MCS

$R_{predict}(\bullet)$: predicted approximated reliability based on • obtained from the LSTM-BAT-MCS

$\delta$: number of coordinates in supervectors

$N_{sim}$: number of simulations in each run of MCS

$N_{sim}(S)$: number of simulations of supervector $S$ in each run of BAT-MCS

$N_{run}$: number of runs for each BAT-MCS

$\varepsilon$: error between the estimator and the exact solution

$\sigma$: standard deviation

$N_{term}$: number of time steps in the collected data

$F(\bullet)$: loss function of •

$P_t$: vector included each component and the approximated reliability at time step $t$

## 2.3 Nomenclature

Binary-state network: A network with binary-state arcs only, that is, the state of each arc is either zero or one.

Reliability: The probability that a network is operating successfully.

Supervector: $S = (s_1, s_2, \ldots, s_\delta)$ is a supervector if $S(a_i) = s_i$ and $i = 1, 2, \ldots, \delta$ and $\delta \leq m$.

Connected vector: Vector $X$ is connected if nodes 1 and $n$ are connected in $G(X)$.

Disconnected vector: Vector $X$ is disconnected if nodes 1 and $n$ are disconnected in $G(X)$.

## 2.4 Assumptions

1. Networks have no parallel arcs or loops.
2. Nodes are all completely reliable and connected.

3. Arcs have two states and all states are statistically independent.

## 3. REVIEW OF BAT, MCS, PLSA, BAT-MCS, AND LSTM

The proposed LSTM-BAT-MCS integrates the BAT-MCS [27] to generate a network reliability dataset, and LSTM to train, learn, and predict [40]. The BAT-MCS also implements BAT to find all supervectors [13], MCS [22, 24, 27] to approximate the reliability of the binary-state network based on each feasible supervector, and PLSA [17] to verify the connectivity of each state vector. Hence, the BAT, MCS, PLSA, BAT-MCS, and LSTM are briefly introduced in this section.

### 3.1 BAT

The original BAT was proposed by Yeh to find all $m$-tuple vectors without duplicities in $O(n2^{m+1})$ based on the binary addition operator [11, 13, 26]. For example, all vectors obtained from the BAT in Figure 1 are listed in Table 2, such that

1. $X_{\text{new}}(a_1) = 1$ and $X_{\text{new}}(a_i) = X_{\text{old}}(a_i)$ for $i = 2, 3, \ldots, m$ if $X_{\text{old}}(a_1) = 0$, for example, $X_{\text{old}} = X$ in iterations 1, 3, 5, …, 31 in Table 2.
2. $X_{\text{new}}(a_j) = 0$ for $j = 1, 2, \ldots, i$, $X_{\text{new}}(a_{i+1}) = 1$, and $X_{\text{new}}(a_k) = X_{\text{old}}(a_k)$ for $k = (i+2), (i+3), \ldots, m$ if $X_{\text{old}}(a_1) = \ldots = X_{\text{old}}(a_i) = 1$ and $X_{\text{old}}(a_{i+1}) = 0$, for example, $X_{\text{old}} = X$ in iterations 2, 4, 5, …, 32 in Table 2.

**Table 2.** All vectors obtained from the BAT on Figure 1.

| iteration | $X$ | iteration | $X$ |
|---|---|---|---|
| 1 | (0, 0, 0, 0, 0) | 17 | (0, 0, 0, 0, 1) |
| 2 | (1, 0, 0, 0, 0) | 18 | (1, 0, 0, 0, 1) |
| 3 | (0, 1, 0, 0, 0) | 19 | (0, 1, 0, 0, 1) |
| 4 | (1, 1, 0, 0, 0) | 20 | (1, 1, 0, 0, 1) |
| 5 | (0, 0, 1, 0, 0) | 21 | (0, 0, 1, 0, 1) |
| 6 | (1, 0, 1, 0, 0) | 22 | (1, 0, 1, 0, 1) |
| 7 | (0, 1, 1, 0, 0) | 23 | (0, 1, 1, 0, 1) |
| 8 | (1, 1, 1, 0, 0) | 24 | (1, 1, 1, 0, 1) |
| 9 | (0, 0, 0, 1, 0) | 25 | (0, 0, 0, 1, 1) |
| 10 | (1, 0, 0, 1, 0) | 26 | (1, 0, 0, 1, 1) |
| 11 | (0, 1, 0, 1, 0) | 27 | (0, 1, 0, 1, 1) |
| 12 | (1, 1, 0, 1, 0) | 28 | (1, 1, 0, 1, 1) |
| 13 | (0, 0, 1, 1, 0) | 29 | (0, 0, 1, 1, 1) |
| 14 | (1, 0, 1, 1, 0) | 30 | (1, 0, 1, 1, 1) |
| 15 | (0, 1, 1, 1, 0) | 31 | (0, 1, 1, 1, 1) |
| 16 | (1, 1, 1, 1, 0) | 32 | (1, 1, 1, 1, 1) |

To obtain the new binary-state vector $X_{\text{new}}$, the above two steps use the addition operator to add (1, 0, …, 0) to the binary-state vector $X_{\text{old}}$. The corresponding BAT source code can be downloaded in [42]. The general BAT pseudocode is described in [11, 13, 26, 27], and listed below:

**Algorithm: BAT**

**Input:** A positive integer $m$.

**Output:** All $2^{|m|}$ $m$-tuple binary-state vectors $X$.

**STEP B0.** Initialize the coordinate index $i = 1$ and the $m$-tuple binary-state vector $X = (0, 0, \dots, 0)$.

**STEP B1.** If $X(a_i) = 0$, let $X(a_i) = 1$, $i = 1$, we have a new $X$, and go to STEP B1.

**STEP B2.** If $i < m$, let $X(a_i) = 0$, $i = i + 1$, and go to STEP B1.

BAT is very simply structured, as there are only three statements in its pseudocode. There are $2^m$ $m$-tuple binary-state vectors that can be found using BAT in $O(2^{m+1})$ [43]. Thus, BAT is very efficient in finding all binary-state vectors. Because only the binary-state vector $X$ is updated repeatedly and no other vector is used, BAT is easy to code, flexible to fit for any problem, and efficient in both runtime and space.

Moreover, according to extensive experiments [11, 13, 26, 27], BAT outperforms related methods such as DFS [32, 33], BFS [16, 31], and UGFM [34]. As a result, many different extensions of BAT have been applied to numerous real-life applications [11, 13, 26, 27, 43, 44, 45, 46, 47].

### 3.2 PLSA

In general, it is always necessary to find all related feasible vectors to find exact solutions to #P-hard problems. A calculation of network reliability is a #P-hard problem, as it requires finding all connected binary-state vectors between nodes 1 and $n$ in the binary-state networks.

As mentioned in Section 3.2, BAT can efficiently find all binary-state vectors. However, these vectors still require verification. The PLSA is a common method to achieve this, and is usually integrated into BAT to calculate network reliability.

The PLSA was developed in [26] based on the layered-search algorithm (LSA) [17] to verify whether nodes 1 and $n$ are connected in the subnetwork corresponding to a binary-state vector obtained from BAT. Note that due to its efficiency and simplicity, the LSA has been extended further to GLSA [43] and TLSA [48] to verify whether all nodes in a subset and all-pair nodes are connected.

The pseudocode for the PLSA is presented below [26].

**Algorithm: PLSA**

**Input:** A binary-state vector $X$.

**Output:** The connectivity of $X$ in $G(X)$.

**STEP P0.** Let $L_1 = \{1\}$, $i = 2$, and $L_2 = \varnothing$.

**STEP P1.** Let $L_i = \{ v \in V \mid$ if $X(a) = 1$, one endpoint of $a$ is in $L_{i-1}$ and another is not in $L_{i-1}$ in $G(X)\}$.

**STEP P2.** If $L_i = \varnothing$, $X$ is disconnected and halt.

**STEP P3.** If $n \in L_i$, $X$ is a connected vector and halt.

**STEP P4.** Let $i = i + 1$, $L_i = \varnothing$, and go to STEP P1.

Layer $L_j$ has at least one node for $j = 1, 2, \ldots, (i-1)$, $L_k = \varnothing$ for $k = i, (i+1), \ldots, n$. Each node can be selected once, and there are at most $n$ nodes in $G(X)$. Therefore, the time complexity of the PLSA is only $O(n)$. The procedure to verify the connectivity of $X = (1, 1, 1, 1, 1)$ using the PLSA is shown in Table 3.

**Table 3.** Example PLSA procedure on $X = (1, 1, 1, 1, 1)$ in Figure 1.

| $i$ | $L_i$ | Remark |
|---|---|---|
| 1 | {1} | |
| 2 | {2, 3} | |
| 3 | {4} | $X$ is connected |

### 3.3 MCS

The MCS is a very useful method to obtain approximate solutions for complex nonlinear problems with high scale and dimension [22, 23, 24, 25, 26, 27], especially NP-hard and #P-hard problems.

To obtain the approximated binary-state network reliability, MCS must generate sufficient random numbers in [0, 1] from the uniform distribution of each arc. Each arc is set to fail if its reliability is greater than its corresponding random numbers. The number of passes increases by 1 if nodes 1 and $n$ remain connected after the removal of failed arcs. The above simulation procedure is repeated for several iterations, and the average number of passes denotes the approximated reliability.

The pseudocode of the BAT-MCS is given as follows [27]:

**Algorithm: MCS**

**Input:** $G(V, E, \mathbf{D})$, the source node 1, the sink node $n$, and the number of simulations $N_{sim}$.

**Output:** The approximated binary-state network reliability $R^*$.

**STEP M0.** Initialize the index of the simulation sim = 0 and the index of the pass number $N_{pass} = 0$.

**STEP M1.** Let $X(a_i) = 1$ if $\rho_i < \Pr(a_i)$ for all $a_i \in E$; otherwise, $X(a_i) = 0$, where $\rho_i$ is a random number generated within [0, 1] uniformly,

**STEP M2.** Let $N_{pass} = N_{pass} + 1$ if nodes 1 and $n$ are connected after using the PLSA in $G(X)$.

**STEP M3.** If sim < $N_{sim}$, let sim = sim + 1 and go to STEP M1.

**STEP M4.** Let $R^* = N_{pass}/N_{sim}$ and halt.

The estimator $R^*$ is unbiased and consistent with $R$ [22, 23, 24, 25, 26, 27]. In addition, the relationships between its relative error ε, confidence interval (1−α)% [22, 23, 24, 25, 26, 27], and $N_{sim}$ can be predicted using the following equation:

$$\frac{Z_{\alpha/2}^2}{4\varepsilon^2} \le N_{sim}. \tag{2}$$

For example, let $N_{sim} = 8$ for the binary-state network shown in Figure 1, with the state distribution listed in Table 1. The MCS procedure is presented in Table 4.

**Table 4.** Example MCS Procedure on Figure 1.

| iteration | $\rho_1$ | $\rho_2$ | $\rho_3$ | $\rho_4$ | $\rho_5$ | $X$ | Connected? | $N_{pass}$ |
|---|---|---|---|---|---|---|---|---|
| 1 | 0.16861 | 0.00281 | 0.48407 | 0.78768 | 0.26430 | (1, 1, 0, 0, 1) | Y | 1 |
| 2 | 0.24745 | 0.57434 | 0.94997 | 0.25461 | 0.81555 | (1, 1, 1, 1, 0) | Y | 2 |
| 3 | 0.37612 | 0.17782 | 0.69576 | 0.25661 | 0.95894 | (1, 1, 1, 1, 0) | Y | 3 |
| 4 | 0.99637 | 0.13337 | 0.93463 | 0.91058 | 0.83461 | (0, 1, 0, 0, 0) | N | |
| 5 | 0.59383 | 0.08648 | 0.52618 | 0.21799 | 0.80758 | (1, 1, 1, 1, 1) | Y | 4 |
| 6 | 0.10244 | 0.41495 | 0.68811 | 0.64513 | 0.44071 | (1, 1, 1, 0, 1) | Y | 5 |
| 7 | 0.00730 | 0.25565 | 0.79960 | 0.82404 | 0.95794 | (1, 1, 0, 0, 0) | N | |
| 8 | 0.46081 | 0.51113 | 0.09275 | 0.93567 | 0.12821 | (1, 1, 1, 0, 0) | N | |
| SUM | | | | | | | | 5/8=0.625 |

### 3.4 BAT-MCS

The BAT-MCS proposed by Yeh [27] combines BAT and MCS. Each vector generated from the BAT component is a special subvector with δ coordinates, called the supervector in [13]. The BAT-MCS finds all supervectors rather than all vectors, making it more efficient than the BAT. After finding each supervector from the BAT, the states of these arcs in the coordinates of the supervectors can be determined. The MCS is applied only to coordinates that are not included in the supervector. In addition, the BAT-MCS implements the self-adaptive algorithm to determine the simulation number.

Moreover, the number of random variables is reduced in the BAT-MCS, which results in more robustness, requires less simulations, and yields more accurate reliability measures than the MCS. Because the BAT-MCS is a powerful tool for estimating binary-state network reliability, it was implemented in our proposed LSTM-BAT-MCS to generate a dataset with different time steps.

**Algorithm: BAT-MCS**

**Input:** $G(V, E, \mathbf{D})$, the source node 1, the sink node $n$, the number of coordinates δ in supervectors, and the number of simulations $N_{sim}$.

**Output:** The approximated binary-state network reliability $R^*$.

**STEP A0.** Let $R^* = 0$.

**STEP A1.** Implement the BAT to find all δ-tuple supervector set Δ.

**STEP A2.** Let $R^* = R^* + \Pr(S)$ and $\Delta = \Delta - \{S\}$ for all $S$ if $X = L(S)$ is connected after using the PLSA, where $X(a_i) = S(a_i)$ for $i = 1, 2, \ldots, \delta$ and $X(a_i) = 0$ for $i = (\delta+1), (\delta+2), \ldots, m$.

**STEP A3.** Let $\Delta = \Delta - \{S\}$ for all $S$ if $X = U(S)$ is disconnected, where $X(a_i) = S(a_i)$ for $i = 1, 2, \ldots, \delta$ and $X(a_i) = 1$ for $i = (\delta+1), (\delta+2), \ldots, m$.

**STEP A4.** Calculate

$$n_{sim}(S) = \left\lfloor N_{sim} \times \frac{\Pr(S)}{\sum_{S\in\Delta} \Pr(S)} \right\rfloor. \quad (3)$$

**STEP A5.** Implement the MCS($N_{sim}(S)$) to have $N_{pass}$ of $G(S)$ and let $R^* = R^* + \Pr(S)\times N_{pass}/N_{sim}(S)$ for all $S \in \Delta$.

The BAT-MCS is more efficient than the BAT if $\delta << m$, and very close to the MCS if $\delta$ is very small from the time complexity. From the computational experiments, the BAT-MCS outperforms the MCS in terms of solution quality and variance, as shown.

For example, let $N_{sim} = 512$ and $\delta = 2$, as shown in Figure 1. The procedure of the BAT-MCS for calculating $R^*$ is presented in Tables 5-7. Based on the BAT, Table 5 lists the complete supervector set $\{S_1, S_2, S_3, S_4\}$, finds $\Delta = \{S_2, S_3, S_4\}$, and calculates $n_{sim}(S)$ for all $S \in \Delta$.

**Table 5.** The procedure of the BAT part in the BAT-MCS.

| $i$ | $S_i$ | $\Pr(S_i(a_1))$ | $\Pr(S_i(a_2))$ | $\Pr(S_i)$ | $n_{sim}(S_i)$ |
|---|---|---|---|---|---|
| 1 | (0, 0) | | | | |
| 2 | (1, 0) | 0.9 | 0.2 | 0.18 | 94 |
| 3 | (0, 1) | 0.1 | 0.8 | 0.08 | 41 |
| 4 | (1, 1) | 0.9 | 0.8 | 0.72 | 376 |
| SUM | | | | 0.98 | 511 |

Based on MCS and $N_{sim}(S)$ for all $S \in \Delta$, Table 6 shows the simulated $N_{sim}(S_i)$ times of the possible cases of $S_i$. For example, for $i = 2$, 0.69576, 0.25661, and 0.95894 were random numbers uniformly generated within [0, 1]. Because 0.69576 < $\Pr(a_3)$ = 0.7, 0.25661 < $\Pr(a_4)$ = 0.6, and 0.95894 < $\Pr(a_5)$ = 0.5, the states of $a_3$, $a_4$, and $a_5$ are all working.

**Table 6.** Procedure of the MCS part in the BAT-MCS.

| $i$ | $S_i$ | $\Pr(S_i)$ | $j$ | $\rho_3$ | $X(a_3)$ | $\rho_4$ | $X(a_4)$ | $\rho_5$ | $X(a_5)$ | $X$ | $X$ is connected? |
|---|---|---|---|---|---|---|---|---|---|---|---|
| 2 | (1, 0) | 0.18 | 1 | 0.69576 | 1 | 0.25661 | 1 | 0.95894 | 0 | (1, 0, 1, 1, 0) | Y |
| | | | 2 | 0.93463 | 0 | 0.91058 | 0 | 0.83461 | 0 | (1, 0, 0, 0, 0) | N |
| | | | : | | | | | | | | |
| | | | 94 | 0.80660 | 0 | 0.24372 | 1 | 0.87651 | 0 | (1, 0, 0, 1, 0) | Y |
| 3 | (0, 1) | 0.08 | 1 | 0.52618 | 1 | 0.21799 | 1 | 0.80758 | 0 | (0, 1, 1, 1, 0) | Y |
| | | | 2 | 0.68811 | 1 | 0.64513 | 0 | 0.44071 | 1 | (0, 1, 1, 0, 1) | Y |
| | | | : | | | | | | | | |
| | | | 41 | 0.70413 | 0 | 0.82262 | 0 | 0.15342 | 1 | (0, 1, 0, 0, 1) | Y |
| 4 | (1, 1) | 0.72 | 1 | 0.79960 | 0 | 0.82404 | 0 | 0.95794 | 0 | (1, 1, 0, 0, 0) | N |
| | | | 2 | 0.09275 | 1 | 0.93567 | 0 | 0.12821 | 1 | (1, 1, 1, 0, 1) | Y |
| | | | : | | | | | | | | |
| | | | 376 | 0.05120 | 1 | 0.77447 | 0 | 0.82728 | 0 | (1, 1, 1, 0, 0) | N |

The corresponding state vector based on $S_2$ and the arc states above is $X = (1, 0, 1, 1, 0)$, where the first two coordinates are from $S_2 = (1, 0)$. As $X = (1, 0, 1, 1, 0)$ is disconnected, the value of $N_{pass}$ remains the same. Similarly, we have the last simulation where $N_{sim}(S_2) = 94$ and all simulations for $S_3$ and $S_4$ in Table 7 for $i = 3$ and 4, respectively.

Assume that the number of passes, that is, $N_{pass}$, based on $S_2$, $S_3$, and $S_4$ is 55, 39, and 237, respectively. Table 7 shows the calculation of $R^*(S_i) = \Pr(S_i) \times N_{pass}/N_{sim}(S_i)$ based on $\Pr(S_i)$, $N_{pass}$, and $N_{sim}(S_i)$, for $i = 2$, 3, and 4. The final $R^*$ is $R^*(S_2) + R^*(S_3) + R^*(S_4) = 0.635246$.

**Table 7.** procedure of the MCS part in the BAT-MCS.

| $i$ | $\Pr(S_i)$ | $N_{pass}$ | $N_{sim}(S_i)$ | $\Pr(S_i)\times N_{pass}/N_{sim}(S_i)=R^*(S_i)$ |
|---|---|---|---|---|
| 2 | 0.18 | 55 | 94 | 0.18×33/94=0.105319 |
| 3 | 0.08 | 39 | 41 | 0.08×39/41=0.076098 |
| 4 | 0.72 | 237 | 376 | 0.72×41/376=0.453830 |
| SUM | | | | 0.635246 |

### 3.5 LSTM

LSTM [40] is a variant of recurrent neural networks (RNN) [41], which is a type of deep learning that specializes in time-series problems, such as natural language processing, time-series classification, and speech recognition. Hence, the input data sequence contains further information such that the outputs of the previous time step are the inputs of the current time step, which is the major difference between LSTM/RNN and other deep learning methods [35].

An LSTM unit is a special ANN consisting of a hidden state as short-term memory and a cell state as long-term memory to carry information along with all the time steps. In LSTM, there are three special gates: the forget gate, the input gate, and the output gate, with vector values $F_t$, $I_t$, and $O_t \in (0, 1)^h$, respectively, where $t$ is the time and $h$ is the number of hidden units. The cell remembers values, and the three gates control the cell's input/output flow over arbitrary time intervals.

Let $\eta$, $\bullet$, $X_t \in R^{\eta}$, $H_t \in (-1, 1)^h$, $\underset{\sim}{C}_t \in (-1, 1)^h$, $C_t \in R^h$, $W_{x,\bullet} \in R^{h\times\eta}$, $W_{h,\bullet} \in R^{h\times h}$, $b_\bullet \in R^h$, $\sigma_s$, and $\sigma_t$ be the number of input features, superscript in $\{f, i, o, c\}$, input vector/matrix, output/hidden state vector, cell input activation vector, cell state vector, weight matrices and bias vector parameters, sigmoid function, and hyperbolic tangent function, respectively, at time $t$.

The forget gate forgets irrelevant information in the output from the previous time step based on the following equation:

$$F_t = \sigma_g(W_{x,f}X_t + W_{h,f}H_{t-1} + b_f). \tag{4}$$

The input gate mines new information from the input according to the following equation:

$$I_t = \sigma_s(W_{x,i}X_t + W_{h,i}H_{t-1} + b_i). \tag{5}$$

The output gate transfers updated information from the current time step to the next time step using the following equation:

$$O_t = \sigma_s(W_{x,o}X_t + W_{h,o}H_{t-1} + b_o) \tag{6}$$

The sigmoid function $\sigma_s$ sets each coordinate value in $F_t$, $I_t$, and $O_t$ to a number between 0 and 1. For example, in $F_t$, 0 means that everything is forgotten whereas 1 means that everything is retained.

The cell state $C_t$ is calculated using the following equation related to the information forgotten from the previous cell state $C_{t\text{-}1}$, i.e., $(F_t{\circ}C_{t-1})$, and the new information mined from the input gate $(I_t{\circ}N_t)$:

$$C_t = F_t{\circ}C_{t-1} + I_t{\circ}N_t \tag{7}$$

where the new information $N_t$ is obtained by the following formula:

$$N_t = \sigma_t(W_{x,c}X_t + W_{h,c}H_{t-1} + b_c). \tag{8}$$

The hidden state $H_t$ is the Hadamard product of the hyperbolic tangent function of the long-term memory ($C_t$) and the current output $O_t$:

$$H_t = O_t{\circ}\sigma_t(C_t) \tag{9}$$

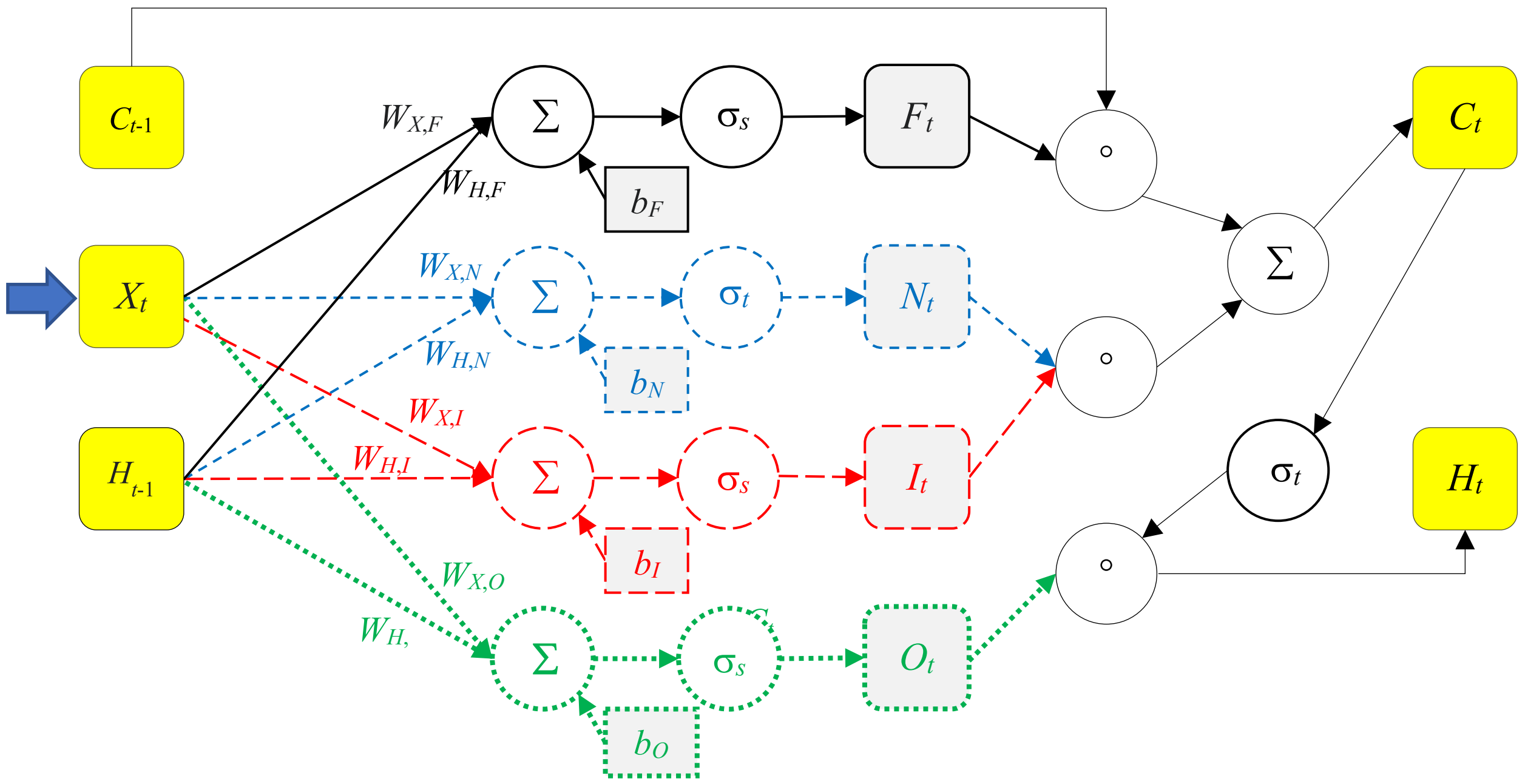

**Figure 3.** Flow information in the LSTM.

The Hadamard product is a point-wise product. The hyperbolic tangent function $\sigma_t$ sets the value of each coordinate in $N_t$ and $H_t$ to a number between −1 and 1. For example, in $N_t$, information subtracted from or added to $C_t$ and $N_t$ is negative or positive, respectively.

Figure 3 illustrates the information flow in the cell layer, hidden layer, and three gates of an LSTM, where each square block, circle, and arrow represent the neural network layer, related operator, and flow direction, respectively.

## 4. PROPOSED LSTM-BAT-MCS

The proposed LSTM-BAT-MCS is split into two components: the binary-state network reliability time-series dataset generated using BAT-MCS, and the LSTM which is used to train, test, and predict the reliability dataset. We address the method of generating the dataset and collecting its features in Section 4.1, and discuss the LSTM components of the proposed algorithm in Section 4.2.

### 4.1 DATA and the BAT-MCS in the LSTM-BAT-MCS

Garbage in, garbage out. Hence, the dataset always plays an important role in LSTM and all machine learning algorithms. Without a good dataset that includes all required information to represent the system we want to predict or analyze, LSTM will yield inaccurate and misleading results. Therefore, the initial step of the proposed algorithm is to collect data to build a reliable time series dataset.

#### 4.1.1 Component Reliability Features

Each component reliability Pr($a$, $t$) is treated as a feature in the dataset for all $a \in E$ and $t \in T$. To meet real-life applications, each component reliability is decreased with a given distribution.

**Table 8.** Example dataset based on Figure 1.

| $t$ | Pr($a_1$, $t$) | Pr($a_2$, $t$) | Pr($a_3$, $t$) | Pr($a_4$, $t$) | Pr($a_5$, $t$) | $R^*(t)$ |
|---|---|---|---|---|---|---|
| 1 | 1.00 | 1.00 | 1.00 | 1.00 | 1.00 | 1.00 |
| 2 | 0.99 | 0.99 | 0.99 | 0.99 | 0.99 | 1.00 |
| | | | | | | |
| 99 | 0.01 | 0.01 | 0.01 | 0.01 | 0.01 | 0.00 |
| 100 | 0.00 | 0.00 | 0.00 | 0.00 | 0.00 | 0.00 |

For example, assume that the value of each component reliability decreases from 1 to 0 within 100 time steps based on the uniform distribution in Figure 1. The dataset includes five features to represent the time-dependent reliabilities of the five arcs in Figure 1, as shown in Table 8.

### 4.1.2 Network Reliability Feature and the BAT-MCS

The approximated reliability $R^*(t)$ is also treated as a feature in the dataset for all $a \in E$ and $t \in T$. To ensure accuracy, the approximated reliability $R^*(t)$ is obtained from the BAT-MCS in terms of Pr($a$, $t$) for all $a \in E$ and $t \in T$. Hence, for a network with $m$ nodes, we have ($m$+1) features in each dataset.

To meet practical applications, the reliability of the binary-state network also varies with time. For example, the last column in Table 1 lists network reliability features $R^*(t)$, based on the values of Pr($a_1$, $t$), Pr($a_2$, $t$), Pr($a_3$, $t$), Pr($a_4$, $t$), and Pr($a_5$, $t$), for $t$ = 1, 2, …, 255. In Table 8, all values of $R^*(t)$ are obtained from the BAT-MCS for $t$ = 1, 2, …, 100, with $||S||$ = 2 for each supervector $S$.

### 4.1.3 Shuffling and Normalizing Data

In deep learning, a dataset is usually cleaned by random shuffling and normalization. Shuffling the data removes possible drifts. However, in each dataset, each component's reliability is decreased and obtained based on its own distribution. In addition, the approximated reliability is estimated from all component reliabilities using the same network configuration based on the BAT-MCS. Therefore, even if the order of $t$ has a different $R^*(t)$, there is no shuffling in the proposed algorithm.

Normalization can increase the cohesion of each feature to have the same range of values, [-0.1, 1.0] in this study. A mean normalization is implemented to normalize each feature in the dataset as follows:

$$x_{\text{new}} = \frac{x_{\text{old}} - x_{\text{average}}}{x_{\max} - x_{\min}} \tag{10}$$

where $x_{\text{new}}$ is the mean normalized value, $x_{\text{old}}$ is the original value, $x_{\text{average}}$ is the sample mean, $x_{\max}$ is the maximum value of all $x$, and $x_{\min}$ is the minimum value of all $x$.

For example, all data in Table 8 are mean normalized in Table 9.

**Table 9.** Example dataset after mean normalizing based on Figure 1.

| $t$ | $\Pr^{\#}(a_1, t)$ | $\Pr^{\#}(a_2, t)$ | $\Pr^{\#}(a_3, t)$ | $\Pr^{\#}(a_4, t)$ | $\Pr^{\#}(a_5, t)$ | $R^{\#}(t)$ |
|---|---|---|---|---|---|---|
| 1 | 0.50 | 0.50 | 0.50 | 0.50 | 0.50 | 0.50 |
| 2 | 0.49 | 0.49 | 0.49 | 0.49 | 0.49 | 0.50 |
| 99 | -0.49 | -0.49 | -0.49 | -0.49 | -0.49 | -0.50 |
| 100 | -0.50 | -0.50 | -0.50 | -0.50 | -0.50 | -0.50 |

### 4.1.4 Training, Validating, and Testing Sets

There are 256 time steps for the data collected from each binary-state benchmark network, and the rolling window of the proposed LSTM-BAT-MCS is 5.

The dataset was split into a training set and a testing set. The former was 90% and the latter was 10%. Each training set was divided into blocks of five time steps or periods: $i$, $(i+1)$, $(i+2)$, $(i+3)$, and $(i+4)$ for training, that is, a rolling window of five periods was used. The next time step for each training set, $(i+5)$, is used for validation. Thus, each series of five time steps is trained to predict the next time step, and 250 blocks were used for training, as shown in Figure 4.

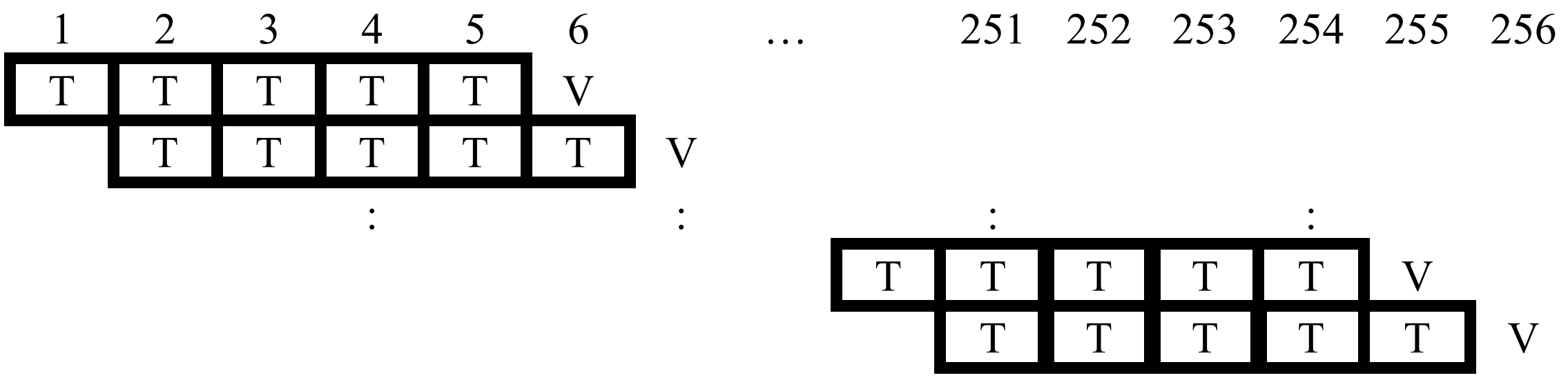

**Figure 4.** Blocks of training (denoted by T) and testing (denoted by V) sets

Each record in the data includes the reliability of each arc and the approximated network reliability obtained from the BAT-MCS. Hence, each record has $|E|+1$ features.

## 4.2 LSTM and the LSTM-BAT-MCS

The LSTM components of the proposed LSTM-BAT-MCS include the loss function, mini-batch gradient descent, adaptive moment estimation (Adam), LSTM network topology, and LSTM-BAT-MCS pseudocode.

### 4.2.1 Loss Function

Let the rolling window in the proposed algorithm be 5. The loss function $F(X_\tau)$, which measures the quality of the parameters, is the mean squared error (MSE) for the $\tau^{\text{th}}$ epoch:

$$F(X_\tau)=\frac{1}{(\mathrm{N}_{\mathrm{term}}-5)}\sum_{t=1}^{(\mathrm{N}_{\mathrm{term}}-5)}[R_{\mathrm{predict},t+5}(P_t\ ,P_{t+1},P_{t+2},P_{t+3},P_{t+4})-R^*(P_{t+5})]^2\ , \tag{11}$$

where $R^*(P_{t+5})$ is the approximated binary-state network reliability obtained from the BAT-MCS, $R_{\mathrm{predict},t+5}(P_t\ ,P_{t+1},P_{t+2},P_{t+3},P_{t+4})$ is the predicted approximated binary-state network reliability in terms of $P_t\ ,P_{t+1},P_{t+2},P_{t+3},P_{t+4}$ , and $\mathrm{N}_{\mathrm{term}}$ is the number of time steps in the collected data.

### 4.2.2 Mini-batch Gradient Descent

Gradient descent is the most common algorithm for optimizing a loss function (for example, Eq. (11)) in deep learning and ANNs [49].

Let $\eta \geq 0$ denote the learning rate. Depending on the number of parameters used to optimize the loss function, there are three variants of gradient descent: batch gradient descent, stochastic gradient descent (SGD) [50], and mini-batch gradient descent [51]. Their formulas are listed in sequence as follows to update all parameters, a single parameter, and specific parameters:

$$X_{\tau+1}=X_\tau-\eta[\nabla_{X_\tau}F(X_\tau)] \tag{12}$$

$$X_{\tau+1}=X_\tau-\eta[\nabla_{X_\tau}F(X_{\tau,t})] \tag{13}$$

$$X_{\tau+1}=X_\tau-\eta[\nabla_{X_\tau}F(X_{\tau,t:t+b})]\ . \tag{14}$$

The use of more parameters yields higher accuracy but requires more runtime [51]. A mini-batch gradient descent method was implemented in this study.

### 4.2.3 Gradient descent Optimizer and Adam

We used Adam as the optimizer for training the proposed LSTM-BAT-MCS. Adam is a widely used gradient descent optimization algorithm based on adaptive estimation of the first-order moment (mean) $m_\tau$ and second-order moment (uncentered variance) $v_\tau$ [52]. Adam is also a replacement optimization algorithm for stochastic objective functions.

As Adam is simple to implement, efficient when running with little memory, and suitable for large-scale problems, it is adapted in our proposed algorithm, and Eq. (14) has been revised accordingly:

$$X_{\tau+1} = X_\tau - \eta \frac{\widehat{m}_\tau}{\sqrt{\widehat{v}_\tau} + \varepsilon} \tag{15}$$

where

$$\widehat{m}_\tau = \frac{m_\tau}{1-\beta_1^\tau} \tag{16}$$

$$\widehat{v}_\tau = \frac{v_\tau}{1-\beta_2^\tau} \tag{17}$$

$$m_\tau = \beta_1 m_{\tau-1} + (1-\beta_1) g_\tau \tag{18}$$

$$v_\tau = \beta_2 v_{\tau-1} + (1-\beta_2) g_\tau^2 . \tag{19}$$

The default values of $\eta$, $\beta_1$, $\beta_2$, and $\varepsilon_1$ are 0.001, 0.9, 0.999, and $10^{-8}$, respectively, in [52].

### 4.2.4 LSTM-BAT-MCS Pseudocode

The LSTM model in the proposed LSTM-BAT-MCS consists of an ($m$+1) input neuron with five time steps, one hidden layer with 10 neurons, and one output neuron. The pseudocode for approximating and predicting the time-dependent network binary-state reliability by integrating the BAT-MCS and LSTM is presented here.

**Algorithm: LSTM-BAT-MCS**

**Input:** A binary-state network $G(V, E, \mathbf{D})$ and the reliability degradation function for each arc, where nodes 1 and $n$ are the source and sink, respectively.

**Output:** Approximate time-dependent reliability $R^*$ under predefined values $\delta$, Pr($a$) for all $a \in E$, and $N_{sim}$.

**STEP 0.** Implement the BAT-MCS to generate dataset.

**STEP 1.** Implement the LSTM to generate dataset.

## 5 EXPERIMENTAL RESULTS

To demonstrate the performance of the proposed LSTM-BAT-MCS, a complete computational experiment on three large-scale binary-state networks is presented here.

## 5.1 Datasets

Three datasets from three tested larger-scale binary networks were used in the experiments, as shown in Figure 5. Their corresponding information is listed in Table 10.

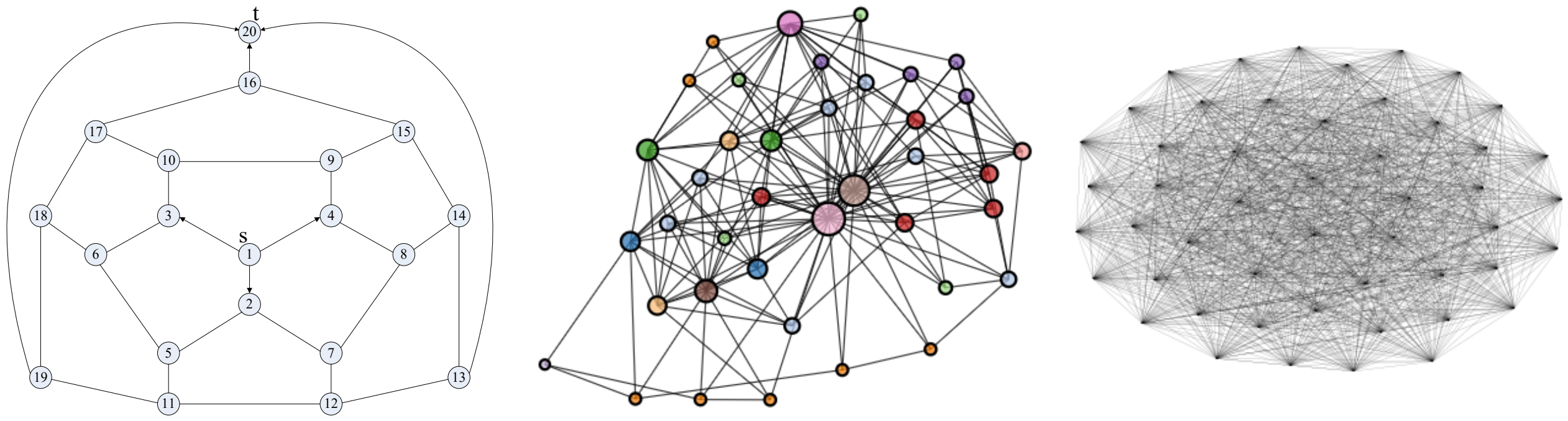

(1) (2) DM-LC (3) $K_{50}$

**Figure 5.** Benchmark networks in experiments.

**Table 10.** Information of all networks in Figure 5.

| $t$ | Figure 5(1) | Figure 5(2) | Figure 5(3) |
|---|---|---|---|
| $\|V\|$ | 20 | 39 | 50 |
| $\|E\|$ | 30 | 170 | 1225 |
| Pr($a$, $t$) | Pr($a$, $t$–1) – 1/(2×256) | Pr($a$, $t$–1)×exp(-1/100) | Pr($a$, 0)/[1+$t$×Pr($a$, 0)] |

### 5.1.1 Three Test Networks

Figure 5(1), with 20 nodes and 30 arcs, is a well-known binary-state network in the research area of traditional network reliability, and is generally utilized to certify the performance of new algorithms in network reliability studies [11]. Among the three networks in Figure 5, only the first benchmark network is adapted from reliability because most benchmark networks are too small to demonstrate the algorithm's actual performance. Even with only 30 arcs in Figure 5(1), it is still very difficult to calculate reliability even for a fixed time step.

Figure 5(2), with 39 nodes and 170 arcs, is a well-known and practical biological network called the DM-LC [53]. These benchmark networks for traditional network reliability problems may be too small, and the number of arcs may be less than 30. Because the DM-LC contains 170 arcs, we used it to test our proposed algorithm.

Figure 5(3), with 50 nodes and 1225 arcs, is the most challenging network in the experiment. This is a complete graph, which implies that any two nodes are connected by one arc [54]. This property makes it more time-consuming to simulate the network's reliability.

### 5.1.2 The Time-Dependent Arc Reliability Functions

To explain and illustrate the effectiveness and importance of the arc reliability function with varying time steps, three different common decay functions were adapted from the zero-order, first-order, and second-order integrated rated laws [55], respectively:

$$\Pr(a, t) = \Pr(a, 0) - t/(2\times 256), \tag{20}$$

$$\Pr(a, t) = \Pr(a, 0)\times\exp(-t/100),\ \text{i.e.,}\ \frac{d\Pr(a,t)}{dt} = -\frac{\Pr(a,t)}{100} \tag{21}$$

$$\Pr(a, t) = \Pr(a, 0)/[1+t\times\Pr(a, 0)],\ \text{i.e.,}\ \frac{d\Pr(a,t)}{dt} = -[\Pr(a,t)]^2 \tag{22}$$

These formulas are also known as exponential decay formulas. Eq. (20) shows that the arc reliability function uniformly decreases with the increment of the time steps, because the right-hand side of Eq. (20) must be subtracted from the constant 1/(2×256). Note that 256 is the number of time steps, that is, $N_{term}$.

The arc reliability function in Eq. (21) decreases exponentially because the right-hand side of Eq. (21) has an exponential base of −1/100. Note that the plot of ln[Pr($a$, $t$)] versus $t$ is a line with a slope of -1/100 and an intercept of ln[Pr($a$, 0)], as shown in Eq. (21). Eq. (22) is complex because the related arc reliability function Pr($a$, $t$) decreases according to the square of itself.

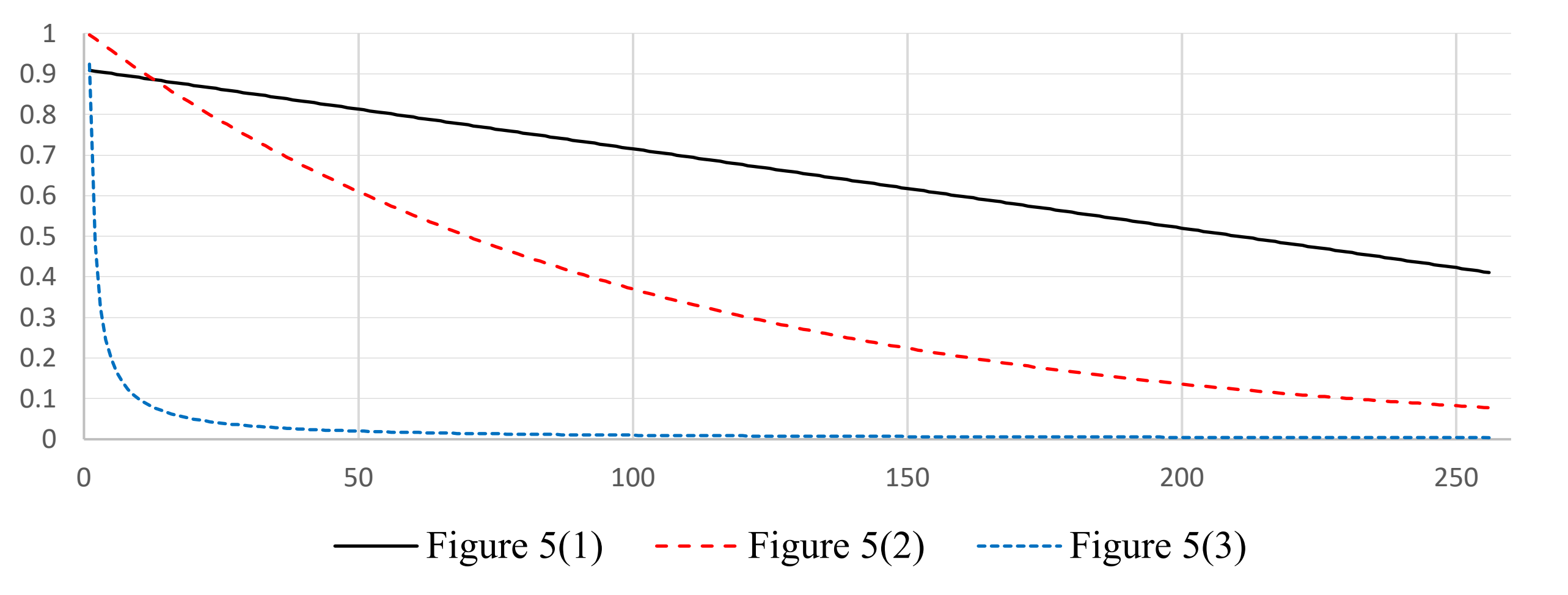

**Figure 6.** The plots for $a_1$ for all $t$ in three test networks

For each test network, the initial reliability of each arc Pr($a$, 0) is between 0.9 and 1.0 uniformly and randomly. Pr($a$, 0) and Pr($a$, $t$) can be obtained using Eq. (20)–(22) for all $a \in E$ and $t \in T$ in Figures 5(1)–(3), respectively. For example, Figure 6 displays the plots for $a_1$ for all $t$ in three test networks.

Figure 6 shows that the curve for Figures 5(1), 5(2), and 5(3) decreases linearly, exponentially, and dramatically, respectively. These trends are based on their respective time-dependent reliability functions.

### 5.1.3 Time-Dependent Approximated Network Reliability

After obtaining the exact time-dependent reliability function of each arc in the three networks shown in Figure 5, the next step is to implement the BAT-MCS component of the proposed algorithm to estimate the approximated time-dependent network reliability.

In the BAT-MCS, $N_{run} = 30$, $N_{sim} = 2^{20}$, $||S|| = n$, and the approximated reliability is the average of 30 approximated reliabilities in each run. Figure 7 depicts the plots of $R^*(t)$ for $t = 1, 2, \ldots, N_{term} = 256$ in the three test networks. Figure 7 shows that the curve for the approximated reliability functions in Figures 5(1) and 5(2) gradually decreases. However, similar to the arc reliability function in Figure 6, the approximated reliability function in Figure 5(3) also decreases dramatically.

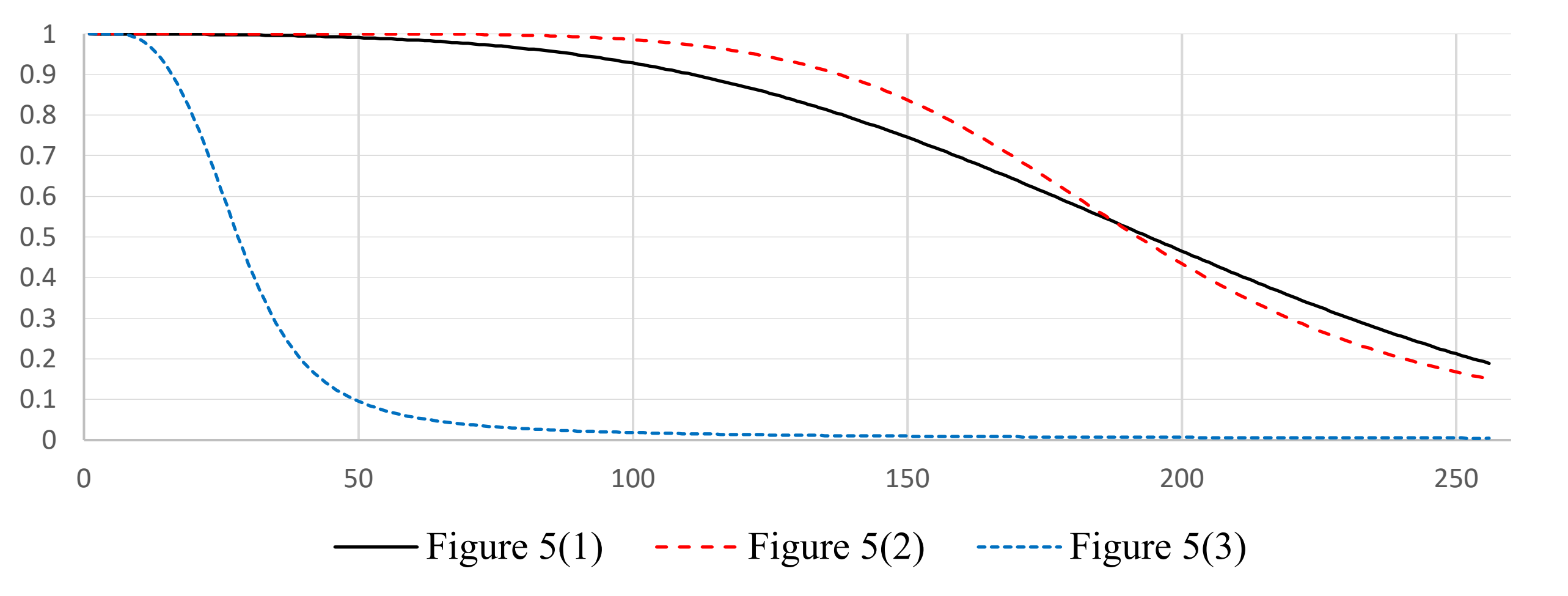

**Figure 7.** The plots for $R^*(t)$ in three test networks.

Note that the exact time-dependent arc reliability function can be given, calculated, or observed; however, network reliability is too difficult to calculate precisely, even for a network with $|E| < 30$.

### 5.1.4 The Split datasets

The dataset is obtained using the exact arc reliability Pr($a$, $t$) and the approximated network reliability $R^*(t)$ obtained from the BAT-MCS for $t = 1, 2, \ldots, N_{term} = 256$. For all networks in the experiments, 90% of each dataset was allocated for training and validation, and the remaining 10% was used for testing. The original and new sizes of each dataset after allocation are listed in Table 11.

**Table 11.** The dataset size in training and testing.

| Dataset size | Figure 5(1) | Figure 5(2) | Figure 5(3) |
|---|---|---|---|
| Origin | 256×31 | 256×171 | 256×1226 |
| LSTM | 250×5×31 | 250×5×171 | 250×5×1226 |
| Train set X | 225×5×31 | 225×5×171 | 225×5×1226 |
| Validating set Y | 225×1 | 225×1 | 225×1 |
| Test set X | 25×5×31 | 25×5×171 | 25×5×1226 |
| Y | 25×1 | 25×1 | 25×1 |

For example, the size of the dataset in Figure 5(1) is 256×31, because Figure 5(1) has $|E| = 30$ arcs, $N_{term} = 256$ time steps, and the approximated network reliability as an additional feature.

The dataset size used in the LSTM was changed from 256×31 to 250×5×31, where 250 corresponds to the block, as in Section 4.1.4. The reason we have 5 in 250×5×31 is that the rolling window is 5 and 256 is reduced to 250, as also explained in Section 4.1.4.

The size of the training set X in Figure 5(1) is 225×5×31, because the dataset used in the proposed algorithm is reduced from 256 to 250 and 90% the data values are 225. Similarly, the size of the validating set Y is 225×1 because the rolling window is 5 and the proposed algorithm is a many-to-one model, where five instances are used to predict one result.

In addition, in Figure 5(1), the size of set X is 25×5×31, and that of set Y is 25×1, because 10% of the 250 instances used in the LSTM are 25.

### 5.2 Results

The proposed LSTM-BAT-MCS was coded in TensorFlow 2.6 with Python 3.7 on a 64-bit Windows 11 Enterprise system with an Intel Core i7-10750H CPU @2.60 GHz and 64 GB RAM.

The results for the loss functions, that is, the MSE shown in Eq.(11), of Figures 5(1), 5(2), and 5(3), are listed in Table 12 and depicted in Figure 8.

**Table 12.** Summary of results.

| items | Figure 5(1) | Figure 5(2) | Figure 5(3) |
|---|---|---|---|
| Compilation Time | 0.00197 | 0.00197 | 0.00199 |
| Training loss | 2.4987e-07 | 7.3177e-06 | 7.3822e-07 |
| Testing loss | 1.6724e-07 | 5.3223e-06 | 6.3380e-07 |
| Total params | 1,680+11 = 1,691 | 7280+11 = 7,291 | 49480+11= 49,491 |

The compilation time was approximately 0.00197, as listed in the first row of each test network in Table 12, regardless of the size of the test networks or the number of parameters used in the proposed algorithm. Therefore, the proposed algorithm was robust.

As shown in Table 12, all final training and testing losses for the three tested networks were extremely close to zero. In addition, as shown in Figure 8, the MSE decreased to almost zero with the epoch number, where Figures 8(1)-8(3) denote the MSE trends for the networks shown in Figures 5(1)-5(3), respectively. Therefore, the proposed LSTM-BAT-MCS can predict the time-independent approximated reliability of the three networks with high accuracy.

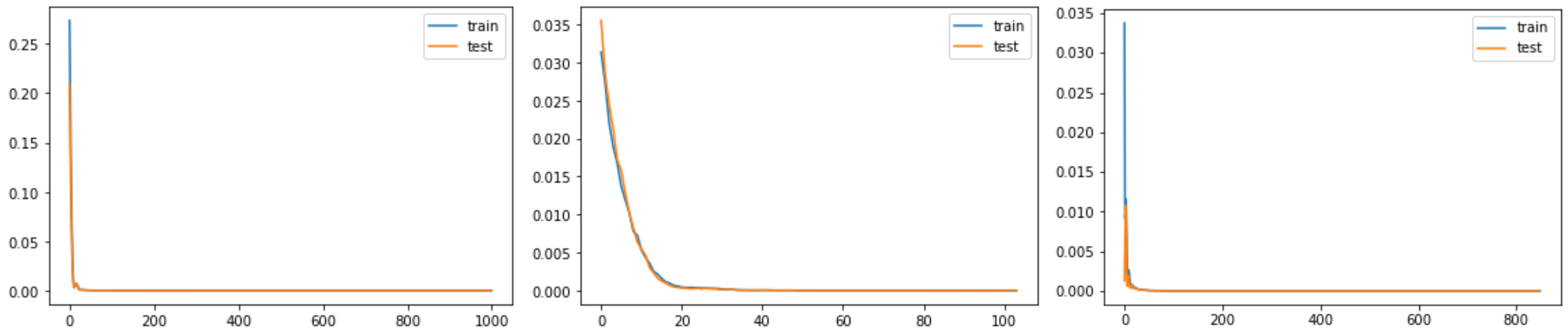

(1) (2) (3)

**Figure 8.** The trends of the MSE values for the three networks.

The last row in Table 12 lists the number of parameters used in the proposed algorithms. The numbers of parameters of the first and second LSTM layers are based on the following equations, respectively:

$$\text{Total parameters of the first layer} = 4 \times h \times (m+1), \quad (23)$$

$$\text{Total parameters of the second layer} = (h+1). \quad (24)$$

Thus, the total number of parameters is $4 \times h \times (m+1) + (m+1)$. For example, Figure 5(1) has 30 arcs, and its parameters are $4 \times h \times (m+1) = 4 \times 10 \times (30+1) = 1680$ in the first LSTM, and $(10+1) = 11$ in the second layer, for a total of 1691.

There are other LSTM versions, such as bidirectional [56], peephole [57], stacked [58], and CNN-LSTM [59]. To ensure solution accuracy, these LSTMs have at least double the number of parameters used by our LSTM. However, from the MSE listed in Table 12, the proposed LSTM-BAT-MCS successfully obtains reliable predicted values.

Figure 9 further compares the predicted values from the proposed LSTM-BAT-MCS to the approximated values obtained from the BAT-MCS, where Figures 9(1)-9(3) denote the MSE trends for the networks shown in Figures 5(1)-5(3), respectively. From Figure 9, these values are extremely close, which also shows the superior performance of the proposed algorithm.

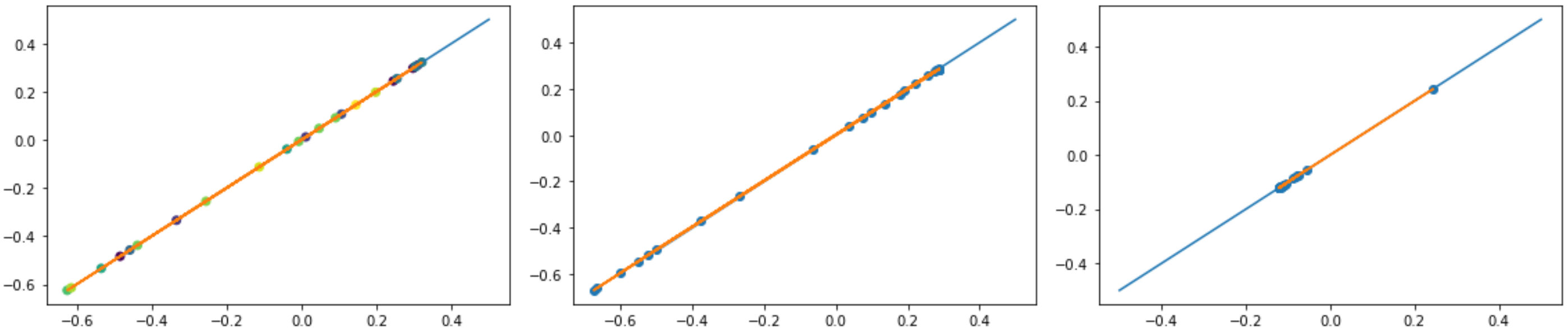

(1) (2) (3)

**Figure 9.** The values obtained from the LSTM-BAT-MCS and BAT-MCS.

The experimental results shown above validate the proposed LSTM-BAT-MCS in terms of the quality of predicted values and the robustness of compilation time.

## 6. CONCLUSIONS

The binary-state network is the basic structure of modern networks, and can be extended to more advanced networks, such as multi-distribution multi-commodity multi-state networks. However, current research focuses on the network reliability problem without considering time as a factor, and research analyzing the best method to approximate and predict time-dependent network reliability, which is more appropriate for practical applications, does not exist.

In this paper, we presented an approach called the LSTM-BAT-MCS for predicting time-dependent binary-state network reliability based on exact time-dependent arc reliability and approximated time-dependent network reliability by integrating the BAT-MCS and an LSTM.

The predicted model was tested on three large networks based on three arc reliability functions. The experimental results validate the performance of the proposed LSTM-BAT-MCS in approximating and predicting time-dependent network reliability.

To the best of our knowledge, this is the first algorithm for time-dependent network reliability.

This study presents a novel LSTM-BAT-MCS by integrating LSTM, BAT, and MCS for the prediction of time-dependent binary-state approximated reliability. In the future, this model will be implemented in real networks to predict reliability at certain time intervals to help decision makers manage the networks. In addition, other deep learning methods such as generative adversarial networks (GANs), incremental learning, and transfer learning will be utilized to enhance the proposed algorithm and apply it to different network structures, such as multi-state networks.

**ACKNOWLEDGMENT**

This research was supported in part by the Ministry of Science and Technology, R.O.C. under grant MOST 110-2221-E-007-107-MY3. This article was once submitted to arXiv as a temporary submission that was just for reference and did not provide the copyright.